\documentclass[twocolumn,showpacs,preprintnumbers,amsmath,amssymb]{revtex4}

\usepackage{dcolumn}
\usepackage{bm}
\usepackage{amsmath,amssymb,amsthm}
\usepackage[dvips]{graphicx}
\usepackage{color}

\def\cald{{\cal D}}
\def\calf{{\cal F}}

\def\nn{\nonumber}
\def\be{\begin{equation}}
\def\ee{\end{equation}}
\def\bea{\begin{eqnarray}}
\def\eea{\end{eqnarray}}
\def\beas{\begin{eqnarray*}}
\def\eeas{\end{eqnarray*}}

\def\l{\left}
\def\r{\right}

\def\bi{\begin{itemize}}
\def\ei{\end{itemize}}
\def\bb{\begin{block}}
\def\eb{\end{block}}

\begin{document}
\title{Delay sober up drunkers: Control of diffusion in random walkers} 

\author{Hiroyasu Ando$^1$,  Kohta Takehara$^2$, Miki U. Kobayashi$^3$}
\affiliation{$^1$Faculty of Engineering, Information and Systems, University of Tsukuba, 1-1-1 Ten-noudai, Tsukuba, 305-8573 Japan,}
\affiliation{$^2$Department of Business Administration,
Graduate School of Social Science, Tokyo Metropolitan University, 1-4-1 Marunouchi, Chiyoda-ku, Tokyo, 100-0005, Japan}
\affiliation{$^3$Faculty of Economics, Rissho University, 4-2-16 Osaki, Shinagawa-ku, Tokyo 141-8602, Japan}

\date{\today}
\pagestyle{plain}

\begin{abstract}
Time delay in general leads to instability in some systems, while a specific feedback with delay 
can control fluctuated motion in nonlinear deterministic systems to a stable state. 
In this paper, we consider a non-stationary stochastic process, i.e., a random walk and observe its diffusion phenomenon with time delayed feedback. Surprisingly, the diffusion coefficient decreases with increasing the delay time.   We analytically illustrate this suppression of diffusion by using stochastic delay differential equations and justify the feasibility of this suppression by applying the time-delay feedback to a molecular dynamics model.  
\end{abstract}

\maketitle

{\em Introduction.---} Diffusion phenomena are ubiquitous in nature, e.g.  in
the Brownian motion \cite{Einstein:1905}, biological membranes \cite{Siwy:2002,Arcizet:2008}, fluid systems \cite{Shraiman:2000}, engineering systems \cite{Collins:1993}, material science \cite{Fujita:2014},  and so on. 
These diffusion phenomena are sources of noise and are unavoidable in experimental systems \cite{Tabard:2007,Butt:1995}.
Noise can prevent the precise manipulation of small scale systems. 
However, it can also be helpful in engineering a system by using 
physical phenomena such as 
Logical Stochastic Resonance, where a certain amount of noise is 
required for logical operation \cite{Murali:2009}. 
In both cases, it is important to control noise in terms of variance of stochastic processes, 
because noise is normally adjusted according to its strength.  
If the variance of noise is controlled, it would be useful for various kinds of applications.

When we consider systems with noise from a theoretical viewpoint, mathematical models  
are convenient, because they can numerically simulate an experiment. 
A mathematical model of the Brownian motion is a typical example of a noisy system,  
which can be modeled by a simple stochastic description, like the Langevin equation. 
For practical purposes, it is important to consider how diffusion can be  suppressed  
in the Brownian motion model. 
The most commonly used approach for suppressing diffusion is to  decrease the temperature of the whole system, resulting in a suppression of thermal fluctuations, 
namely the relation $D\sim~T$, where $D$ is the diffusion coefficient and $T$ is temperature.  
However, the temperature is decreased for all elements in the system and is not selective for specific targets. Therefore, such a control is expensive. 
We introduce an alternative method that can control the diffusion processes of one particle in the Brownian motion.

In terms of cooling nano- microscopic systems, a mathematical model of feedback cooling in electromechanical oscillator \cite{Bonaldi:2009} and a generalized model of Langevin dynamics with non-Markovian feedback \cite{Munakata:2014} have been explored. In these systems, time delay 
in feedback loop is inevitable due to a time lag between the sensor and the manipulator. 
Furthermore, in nonlinear systems, a time-delayed feedback control (DFC) method 
has been proposed to stabilize regular  
motion in fluctuated dynamics \cite{Pyragas:1992}. 
In general, time delay leads to system instability. However, DFC proves that time delay can stabilize deterministic systems. On the other hand, in stochastic systems, e.g. 
a random walk model with time delay, destabilization has been observed as a result of delay  \cite{Ohira:1995}. 
Thus, we consider whether DFC can stabilize non-stationary stochastic systems. 

In this Letter, we focus on stochastic diffusion of random walks,   
corresponding to the Brownian motion in discrete time, 
and apply the DFC method to control the extent of  this diffusion. 
Firstly, we observe diffusion processes in a one-dimensional (1D) random walk model,  
controlled by the DFC method and analyze the control in terms of stochastic delay differential equations. Secondly, we apply the method to a molecular dynamics model of two flocculated particles with thermal fluctuations 
and that interact via the  Lennard-Jones (LJ) potential.  
We numerically confirm that it is possible to suppress diffusion in the molecular dynamics model by the DFC method. 
It is noticed that, in conventional control theory, a single path is controlled to a certain state. In contrast, the purpose of this study was to statistically control  stochastic processes.

{\em Delayed feedback control of random walk.---} In general, nonlinear functions with neutral fixed points show noise-induced diffusion phenomena. 
If these functions are linearized around the neutral fixed points, the following random walk model 
can be derived. 
\begin{equation}
x_{n+1}=x_n+D\xi_n, \label{random}, 
 \end{equation}
 where $x_n$ is a state variable for discrete time $n$,  $\xi_n$ is a $m$-dimensional random variable following the normal distribution $N(0,\ 1)$ for each entry, and 
 $D$ is an amplitude vector of the noise.  Without loss of generality, we consider a 
 1D  case, where $m=1$.
 
 In order to control the random walk in $x_n$, we introduce 
 a time-delayed feedback control method proposed by Pyragas \cite{Pyragas:1992} that is defined by the expression: 
 \begin{eqnarray}
x_{n+1}&=&x_n+D\xi_n - K(x_n-x_{n-\tau}), \label{DFCR} 
 \end{eqnarray}
 In eq. (\ref{DFCR}), the DFC is applied to the model (\ref{random}), where $K$ is   
 feedback gain with a value of $0\le K<1$ and $\tau$ is the delay time. 
 Any nonlinear function with noise can be linearized around its stable fixed point and 
 described in the form of (\ref{DFCR}) as follows. 

If we consider the system:  
 \begin{equation}
 x_{n+1}=f(x_n,a), 
 \end{equation}
 where the map $f$ is nonlinear and has a stable or neutrally stable fixed point. $a$ is a parameter.
We assume that the fixed point is denoted by $\tilde{x}$ and linearize 
$f$ around $\tilde{x}$. This results in:  
 \begin{equation}
 x_{n+1}=f(\tilde{x}) + f'(\tilde{x}) (x_n-\tilde{x}),  
 \end{equation}
 which can be described as: 
 \begin{equation}
 x_{n+1}=f'(\tilde{x}) x_n +(1-f'(\tilde{x}))\tilde{x},   \label{line}
 \end{equation}
 where we use the relation $\tilde{x}=f(\tilde{x})$. 
 For this equation, if we denote $K=1-f'(\tilde{x})$: 
 \begin{equation}
 x_{n+1}=(1-K)x_n+K \tilde{x}. \label{modify}
 \end{equation}
Regarding this linear map, we add the noise term and 
 rewrite (\ref{modify}) as: 
   \begin{equation}
 x_{n+1}=x_n+D \xi_n - K(x_n-\tilde{x}).  \label{no_delay}
 \end{equation}
  The dynamics of the system (\ref{no_delay}) is simple such that a trajectory 
 fluctuates around the fixed point $\tilde{x}$. 
It is possible to suppress diffusion of  a random walk by the form of (\ref{no_delay}), in which 
 the diffusion coefficient is zero.  
 On the other hand, in order to control a non-zero diffusion coefficient, we add  the control term $K(x_{n-\tau}-\tilde{x})$. Consequently, the system 
 is in the form of DFC, namely eq.  (\ref{DFCR}).  

 It should be noted that this discussion is relevant for systems with only stable fixed points and thus, with no diffusion observed. 
 By the procedure above, such stable systems are converted to systems showing diffusion because of the 
 neutral fixed points. 
   
\begin{figure}[htbp]
\begin{center}
\rotatebox{-90}{\includegraphics[width=5.5cm]{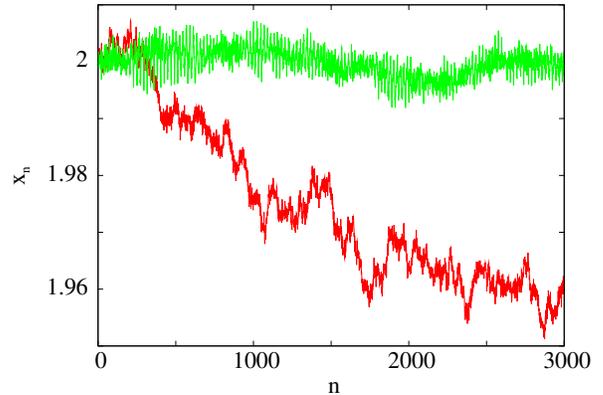}}
\end{center}
\caption{Suppression of diffusion in random walks of the system (\ref{DFCR}). The red solid and green dashed lines represent when $\tau=2$ and 
$\tau=20$, respectively. $D=0.001$.}\label{sup_RW}
\end{figure}

\begin{figure}[htbp]
\begin{center}
\rotatebox{-90}{\includegraphics[width=5.5cm]{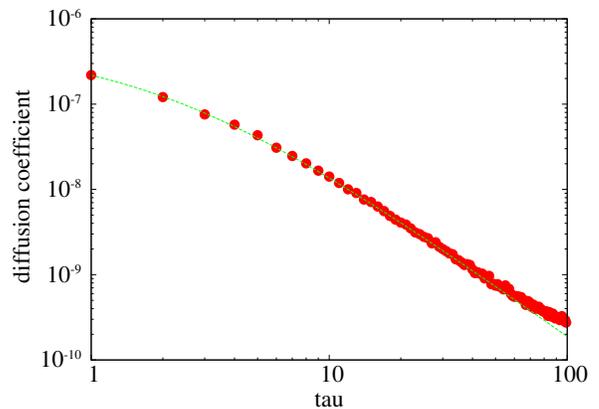}}
\end{center}
\caption{$\cald$ with respect to $\tau$, averaged over $1000$ realizations. The dashed line is proportional to $1/(1+K\tau)^2$. $K=0.5$. }\label{DC}
\end{figure}

Figure \ref{sup_RW} shows the time series of the  (\ref{DFCR}) system dynamics.  
It is observed that a longer $\tau$ suppresses the diffusion of a random walk. 
This result is counterintuitive, since long time delay should  make a system 
unstable. However, in the system (\ref{DFCR}), the opposite is observed. 
Thus, we quantify the diffusion decay  with respect to $\tau$.
In Figure \ref{DC}, the diffusion coefficients $\cald$, defined by $\lim_{T \to T_{\infty}} \langle (X_T - X_0)^2 \rangle/T$, is shown as a function of $\tau$. The result is averaged over $1000$ realizations. 
We takes $T_{\infty}=100000$. It is evident that $\cald$ decays by $\tau^{-2}$. The decay order is illustrated in the following. 

We consider a continuous time (\ref{DFCR}) system: 
\begin{equation}
\dot{y}_t=D\xi_t - K(y_t - y_{t-l}), \label{conti}
\end{equation}
where $y$ is a real variable and $t$ and $t-l$ are continuous time points. 
For this system, let $\Delta=l/\tau$, $\tau \in \{2,3,\ldots\}$. Then, we define $t_n$ as $n\Delta$,  that is $n\frac{l}{\tau}$, where 
$n\in \{-\tau,-\tau+1,\ldots,0,1,\ldots \}$.   If we define $Y_n=y_{t_n}$ and $\xi_n=\xi_{t_n}$,  eq. (\ref{conti}) can 
be rewritten as $Y_{n+1}=Y_n+D\Delta \xi_n - \Delta K (Y_n - Y_{n-\tau})$.  This form is 
equivalent to eq. (\ref{DFCR}).

We consider the continuous time system (\ref{conti}) using stochastic delay differential equations. 
In the standard setting of stochastic calculus, eq. (\ref{conti}) is now redefined as
\be
dy_t = K(y_{t-\tau}-y_t) dt + D dw_t;\ t \geq 0 \label{SDDE0_main}
\ee
where $\{w_t\}_{0\leq t \leq T}$ is a standard Wiener process on a filtered probability space $(\Omega, \calf, \{\calf_t\}, P)$,
and $y_s$ for $s\in [\tau,0]$ is assumed to be 
a given $\calf_0$-measurable random variable $Z_s$. It should be noted that 
$\calf_t$ contains all information for the realization of $w_s$ and $y_s$ for $s\leq t$.
Thus, given $\calf_t$, all of these variables can be all treated as constants.
In the following, the method for determining the diffusion coefficient $\cald$ is briefly described.
For details, see Supplemental Materials.

Equation (\ref{SDDE0_main}) has the unique solution of
\bea
y_t &=& y^Z_t + D \int_0^t y^0_{t-s }dw_s; 0 \leq t, \label{Y_main} \\
y_t &=& Z_t; -\tau \leq t \leq 0 \nn
\eea
where 
\bea
y^Z_t &:=& y^0_t Z_0 + K \int_{-\tau}^{0} y^0_{t-s-\tau} Z_s ds \nn\\
y^0_t &:=& \sum_{n=0}^{\lceil t/\tau \rceil} \l( K^n \over n!\r)
(t- n\tau)^n \exp(-K(t-n\tau)). \label{y0_main}
\eea
(See \cite{Kuchler:1992}).
It should be noted that  $y^0_t$ is a deterministic, rather than a  stochastic process.

Then, we can express $\cald$ in terms of $y^0_t$ as: 
\be
\cald = 
\lim_{T\rightarrow \infty} {D^2 \over T}  \int_0^T (y^0_{s})^2 ds. \label{D_main}
\ee
Moreover, we examined how the function $y^0_t$ behaves in our numerical computations of the equation (\ref{y0_main}),
and found that it seemed converging to a constant of $\bar{y}\propto {1\over 1+ K\tau}$, 
as described in the Supplemental Materials and in \cite{Kobayashi:prep}.
As a consequence, $\cald$ is given by: 
\beas
\cald = (D\bar{y})^2.\nn
\eeas
In addition, it should be noted that this $\bar{y}$ and $\cald$ decreases with increasing $\tau$,
which is consistent with Figure \ref{DC}.

Here, the characteristics of our model should be stressed.
In \cite{Kuchler:1992},  the equivalent conditions to stationary solutions of $y=\{y_t\}$ are given,
which are not met in our system (\ref{Y_main}).
In the case of the stationary $y$, its diffusion coefficient $\cald$ goes to zero as $T \rightarrow \infty$.
Conversely, $\cald$ seems to explode when the solution $y$ is non-stationary.
In this sense, our model with the non-zero $\cald$ lies at the boundary between a stationary and non-stationary state.

This can be seen in another intuitive way.
If a limiting process of our model (\ref{SDDE0_main}) is considered with $\tau \rightarrow 0$: 
\beas
dy^{\downarrow}_t = D dw_t.\nn
\eeas
It should be noted that this is just a scaled random walk, which still includes  $\cald=D$.
On the other hand, 
$\cald = 0$, 
under another limiting process with $\tau \rightarrow \infty$: 
\beas
dy^{\uparrow}_t = K(y_{-\infty} - y^{\uparrow}_t )dt + D dw_t\nn
\eeas
where $y_{-\infty}:=\lim_{t\rightarrow -\infty} y_t$ is given as an initial condition
and as a constant target control level.

As derived before , in our model,  the diffusion coefficient is given by $\cald=D^2\times \l({1\over 1+K\tau} \r)^2$.
Thus, any level of $\cald$ between $(0,D]$ can be achieved with an appropriate choice of $K$ and $\tau$.


{\em Application to Molecular Dynamics. ---} In the following, we apply the control scheme which is shown in the above to a molecular dynamics model.

We consider the dynamics of two particles with noise, and attractive and dissipative forces in a 1D space:
\begin{eqnarray}
  {dx_i \over dt}&=&v_i, \\
  {dv_i \over dt}&=&P_i(\{ x_j\}^2_{j=1})+D_i(\{ v_j\}^2_{j=1})+ R \xi, \label{second}
\end{eqnarray}
where $x_i$ and $v_i$ are the position and velocity of the $i$-th particle ($i=1, 2$),
and $\xi$ is noise from the Gaussian distribution with intensity, $R$.
For simplicity, we assume that the mass of the $i$-th particles, $m_i$=1.
The first term in eq. (\ref{second}) is a conservative force derived as a derivative of the potential $U$ written as: 
\begin{eqnarray}
  P_i &=&-{d U(\{ x_j\}^2_{j=1}) \over dx_i}, \\
  U(\{ x_j\}^2_{j=1}) &=& \sum_{i=1}^{2} \sum_{j=i+1}^2 \tilde{V}(x_{ij}).
\end{eqnarray}
where $\tilde{V}$ is a central force potential, and $x_{ij}=x_i-x_j$.
In this paper, we use the Lennard-Jones potential as the central potential as given by: 
\begin{eqnarray}
  \tilde{V}(x)=\epsilon(({\sigma \over x})^{12} - ({\sigma \over x})^6).
\end{eqnarray}
The second term in eq. (\ref{second}) represents a dissipative force:
\begin{eqnarray}
  D_i=-\gamma v_i,
\end{eqnarray}
where $\gamma$ is a coefficient of drag.
We set $\sigma=2^{-1/6}$ so as that the LJ potential $\tilde{V}(x)$ has a minimum at $x=1$ and $\epsilon=1$.

\begin{figure}
\begin{center}
\includegraphics[width=8.4cm]{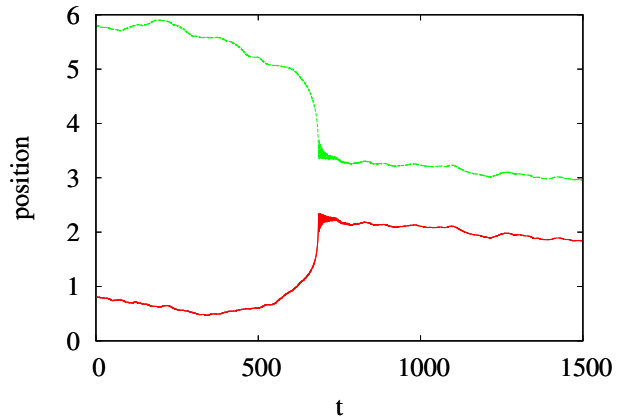}   
\caption{Time evolutions of the positions of two particles without control inputs. 
Particles flocculate because of the interacting force, and move together diffusively.
\label{fig:diffusion}}
\end{center}
\end{figure}

We calculate dynamics of two particles with initial conditions $x_1=0.8, v_1=0.0, x_2=5.8$, and $v_2=0.0$.
In this case, the particles flocculate becase of the attracting force from the LJ potential, and move together diffusively (see Fig. \ref{fig:diffusion}).

\begin{figure}
\vspace{-15mm}
\begin{center}
\includegraphics[width=8.4cm]{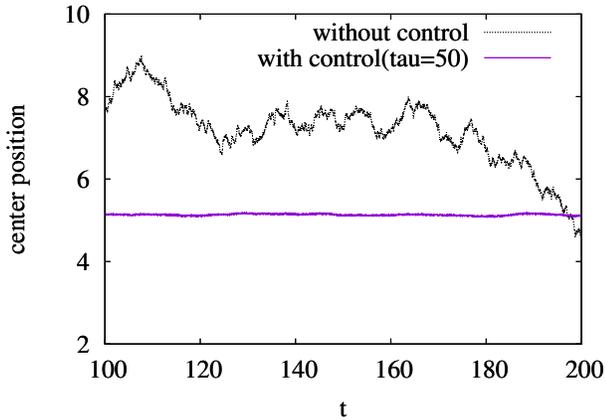}  
\caption{Time evolutions of the center position of two particles. The green and the red lines are the time evolutions without control inputs and with control input ($\tau=50$) , respectively.} 
\label{fig:particle-control}
\end{center}
\end{figure}

In this application, we control fluctuations in the system by using the time-delayed feedback control method.
Velocities fluctuate randomly around $0$ because of the Gaussian noise and are stable at 0 without the noise.
Thus, the dynamics of the velocities is the same as the Gaussian noise.
This Gaussian noise-like behavior  makes the dynamics of the positions analogous to a simple random walk, $ {dx_i \over dt} \sim \xi$. 
Thus, control inputs are given only according to the dynamics of the positions, which are driven by noise originally from the velocities.
The control scheme for the molecular dynamics model is written as:
\begin{eqnarray}
  {dx_i \over dt}&=&v_i-K(x_i-x_i(t-\tau)), \\
  {dv_i \over dt}&=&P_i(\{ x_j\}^2_{j=1})+D_i(\{ v_j\}^2_{j=1})+ R \xi. 
 \end{eqnarray}
Control signals are injected  when the  particles are flocculated. Figure \ref{fig:particle-control} shows the dynamics of the center positions of two particles without control (green dotted line) and with control with $\tau=10$ (red solid line).
The movement of the flocculated particles is diffusive because of noise.
The figure shows that the diffusion of the flocculated particles is suppressed by our control method.

\begin{figure}
\begin{center}
\rotatebox{-90}{\includegraphics[width=6cm]{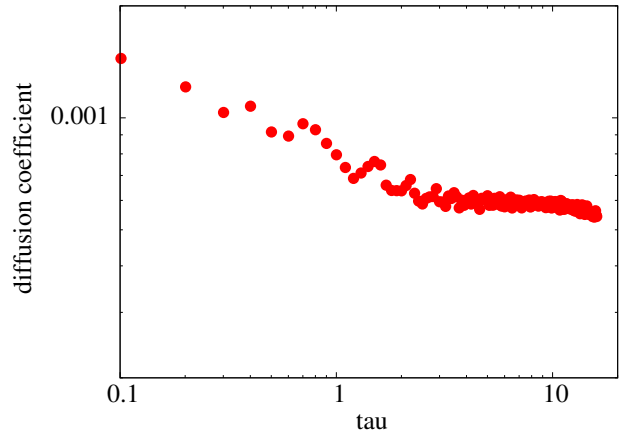}}    
\caption{The diffusion coefficient $\cald$ of the molecular dynamics model as a function of  $\tau$. } 
\label{fig:tau-vs-diff}
\end{center}
\end{figure}

Figure \ref{fig:tau-vs-diff} shows the dependence of the $\cald$ with  $\tau$.
As is the case with the simple random walk, 
the $\cald$ decreases with a change in $\tau$.

{\em Summary.---} In this work, we have applied  time-delayed feedback control 
to a 1D random walk model.  
The interaction between noise and time delay is non-trivial.
We have counterintuitively observed that the diffusion coefficient decreases with increasing delay time. Thus, we have analytically explained the decay of  the diffusion coefficient by solving stochastic delay differential equations derived from the controlled system. In principle, it is possible to control the amount of diffusion from zero to a specific value. 

We have applied the proposed method to a molecular dynamics model described by 
continuous time differential equations. We have numerically demonstrated the control of 
two flocculated particles with the LJ potential. 
Consequently, diffusion generated by noise was successfully controlled and  
decay of the diffusion coefficient was observed in this system.   

The proposed method does not change physical properties of the system in comparison with the conventional technique of reducing diffusion, i.e.  decreasing 
temperature, which can change the thermophysical properties. 
Moreover, the previous analysis of delayed random walks is based on the assumption of a stationarity  \cite{Ohira:1995,Ohira:2000}. In contrast, 
our method is non-stationary and  a conventional analysis by stochastic delay differential equations in \cite{Kuchler:1992} cannot be applied. 
In addition to such theoretical findings, we remark that 
the method can be applied to 
nano- microscopic resonators for reducing the thermal noise. 
In those experimental systems, time delay cannot be avoided or shortened less than some length \cite{Bonaldi:2009,Munakata:2014}. However, making longer time delay can be easy so that our method is effective in the systems.

Other types of time-delayed feedback control are also feasible. For example, we have proposed the adaptive time delay system as follows \cite{Ando:2015}: 
$x_{n+1}=x_n+D\xi_n-1/x_{n-\tau}(x_n-x_{n-\tau})$. This form of adaptive time-delayed feedback control shows almost similar performance to the original system (\ref{DFCR})
in terms of control of the diffusion coefficient. The difference between the original and 
the adaptive control is in determining $K$ or determining the initial values for 
the system. We can apply the adaptive control scheme to real world systems depending on 
the availability of the control parameters. If we cannot access $K$, adaptive feedback gain 
$1/x_{n-\tau}$ can be alternatively used.

\end{document}